\documentstyle[11pt,newpasp,twoside,epsfig]{article}
\def\edcomment#1{\iffalse\marginpar{\raggedright\sl#1\/}\else\relax\fi}
\reversemarginpar

\begin{document}
\title{Nonthermal Activity and Particle Acceleration in Clusters of Galaxies}
\author{Vahe' Petrosian}
\affil{Stanford University, Stanford, CA 94305-4060}
%
%\affil{The Name of My Institution, The Full Address of My Institution}

\begin{abstract}
Evidence for nonthermal activity in clusters of galaxies is well established
from radio observations of synchrotron emission by relativistic electrons, and
new windows (in EUV and Hard X-ray ranges) have provided more powerful tools
for its investigation.  The hard X-ray observations, notably from Coma, are
summarized and results of a new RXTE observations of a high redshift cluster
are presented.  It is shown that the most likely emission mechanisms for these
radiations is the inverse Compton scattering of the cosmic microwave background
photons by the same electrons responsible for the radio radiation.  Various
scenarios for acceleration of the electrons are considered and it is shown that
the most likely model is episodic acceleration by shocks or turbulence,
presumably induced by merger activity, of high energy electrons injected into
the intercluster medium by galaxies or active galactic nuclei.  
\end{abstract}

\section{Introduction}

The intercluster media (ICM) of several clusters of galaxies, in addition to
the well studied thermal bremssstrahlung (TB) emission in the 2 to 10 keV soft
X-ray (SXR) region, show growing evidence for nonthermal activity, first
observed in form of diffuse radio radiation (classified either as relic or halo
sources) and more recently, at extreme ultraviolet (0.07-0.4 keV; EUV) and hard
X-ray (20 to 80 keV; HXR) regions.  In the next section I will give a brief
review of the status of these observations and present new yet unpublished HXR
observation of another cluster, and in \S 3 I will compare the merits of
different emission mechanisms proposed for production of these radiation.  Even
though the presence of nonthermal electrons in the ICM was established decades
ago very little theoretical treatment of the acceleration mechanism was carried
out (see e.g.  Schlikeiser, Siervers \& Thiemann 1987) until the discovery of
the EUV and HXR radiations.  Given the meager amount of the data, detailed
calculations of the energy sources and the exact mechanisms of the acceleration
may be premature.  Consequently, I will emphasize the general physical
characteristics and not the numerical details of the problem.  It turns out
that one can put significant and meaningful constraints on the general aspects
of the acceleration mechanism.  I will describe these in \S 4.

\section{Observations}

The first cluster observed to have {\bf diffuse radio emission} was the Coma
cluster and recent systematic searches have identified more than 30 cluster
with halo or relic sources.  The rate of occurrence of these sources increases
with cluster redshift $z$, SXR luminosity or temperature $T$ (Giovannini \&
Feretti 2000).  There is little doubt that this radiation is due to synchrotron
emission in a magnetic field of strength $B\sim \mu{\rm G}$ by a population of
relativistic electrons of Lorentz factor $\gamma \sim 10^4$.  In the case of
Coma the electron spectra may be represented by a broken power law (Rephaeli
1979) or a power law with an exponential cutoff (Schlikeiser et al.  1987).

{\bf Extreme UV} (0.07 to 0.4 keV) radiation was detected by the {\it Extreme
Ultraviolet Explorer} from Coma (Liu et al.  1966) and some other clusters.  A
cooler ($kT\sim 2keV$) component and inverse Compton (IC) scattering of cosmic
microwave background (CMB) photons by relativistic ($\gamma \sim 10^3$)
electrons are two possible ways of producing this excess radiation.  Some of
the observations and the emission process are still controversial (see ASP
Proc..  301, 2003, eds.  Stewart Bowyer \& Chorng-Yuan Hwang).  I will not
discuss this emission any further here.

The third evidence for nonthermal activity comes from the observations of
excess {\bf HXR} emission in the 20 to 80 keV range by instruments on board
{\it Beppo}SAX and {\it RXTE} satellites.  Each of these observatories has
detected HXR excess from Abell 2256 once and Coma twice, although the second
{\it Beppo}SAX observation shows a weaker signal (Fusco-Femiano et al.  1999
and Rephaeli et al.  1999, 2002).  HXR detections have also been reported in
Abell clusters 754, 2199, 2319 and 3667 all in the redshift range
$0.023<z<0.056$.  Most of these excesses can be best fitted with a fairly hard
spectrum (photon power-law index $\alpha \sim 2$).  Recently, detection of
nonthermal X-rays (albeit at lower energies) have been reported from a
\underline{poor} cluster IC 1262 (Hudson et al.  2003).  In Figure 1 I show the
HXR spectrum and its characteristics from cluster RX-J0658 with a considerably
higher redshift ($z=0.296$) obtained by {\it RXTE}.  These observations
encompass a wide range of temperature, redshift and luminosity, indicating that
HXR emission may be a common property of all clusters with significant diffuse
radio emission.

\begin{figure}[htbp] 
\leavevmode\centering 
\psfig{file=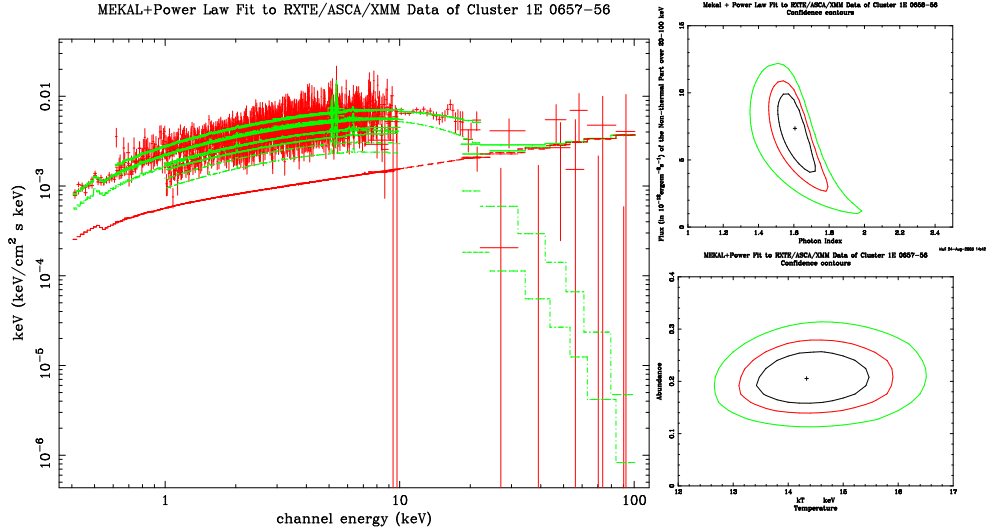,width=70mm,height=130mm, angle=270}
\caption{\footnotesize {\bf Left panel:}  Thermal plus a power law fit to 300ks
RXTE and ASCA+XMM observations of the cluster RX J0658-5557, and the 68, 90 and 
99 \% confidence levels of the photon power-law index vs. 20 - 100 keV 
flux  and temperature vs hydrogen column density.  From Petrosian, Madejski and 
Luli, in preparation.  
} 
\end{figure}

Figure 2 (left panel) shows the photon flux at all wavelengths from Coma, where
in addition to the above mentioned radiations, we show the gamma-ray upper
limit from EGRET on board {\it CGRO} (Sreekumar et al.  1996), and the
equivalent flux for the CMB and optical radiation density present in the
cluster.  (To these should be added the contribution from Far IR background
radiation.)  Similarly, an equivalent flux has been indicated for the static
magnetic field of $\sim 1 \mu$G, which is the size of the field strength
deduced in several clusters (Eilek 1999, Clarke et al.  2001).  The observed
Faraday rotation of the Coma cluster can be interpreted as indicating a uniform
magnetic field along the line of sight of $\sim 0.3\mu$G.  However, the field
lines are most likely chaotic.  Kim et al (1990) and Clarke et al.  (2003)
estimate a mean magnetic field of $\sim 2-3\mu$G.  It should be however noted
that the interpretation of these observations is controversial (see Rudnick \&
Blundell 2002).

\begin{figure}
\plottwo{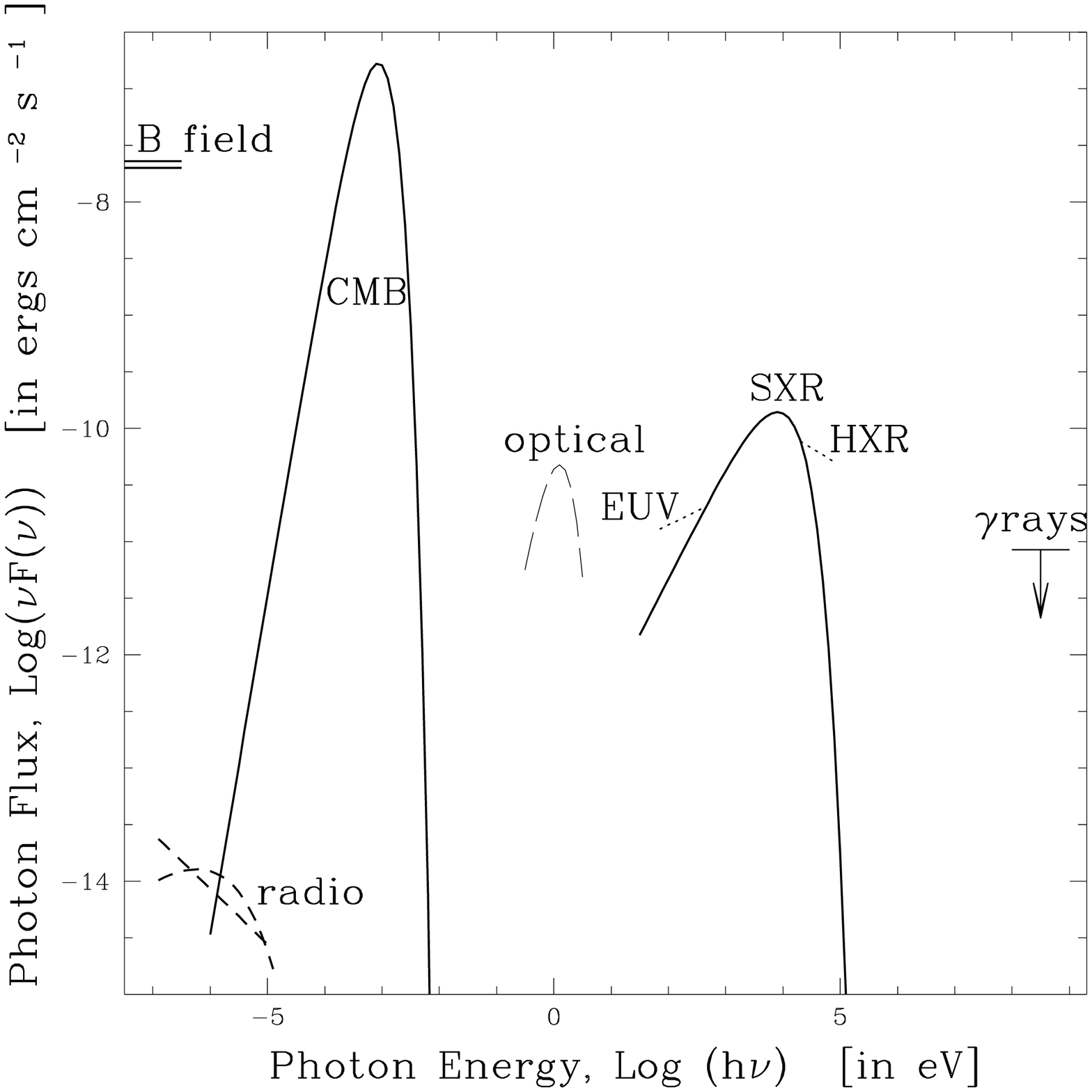}{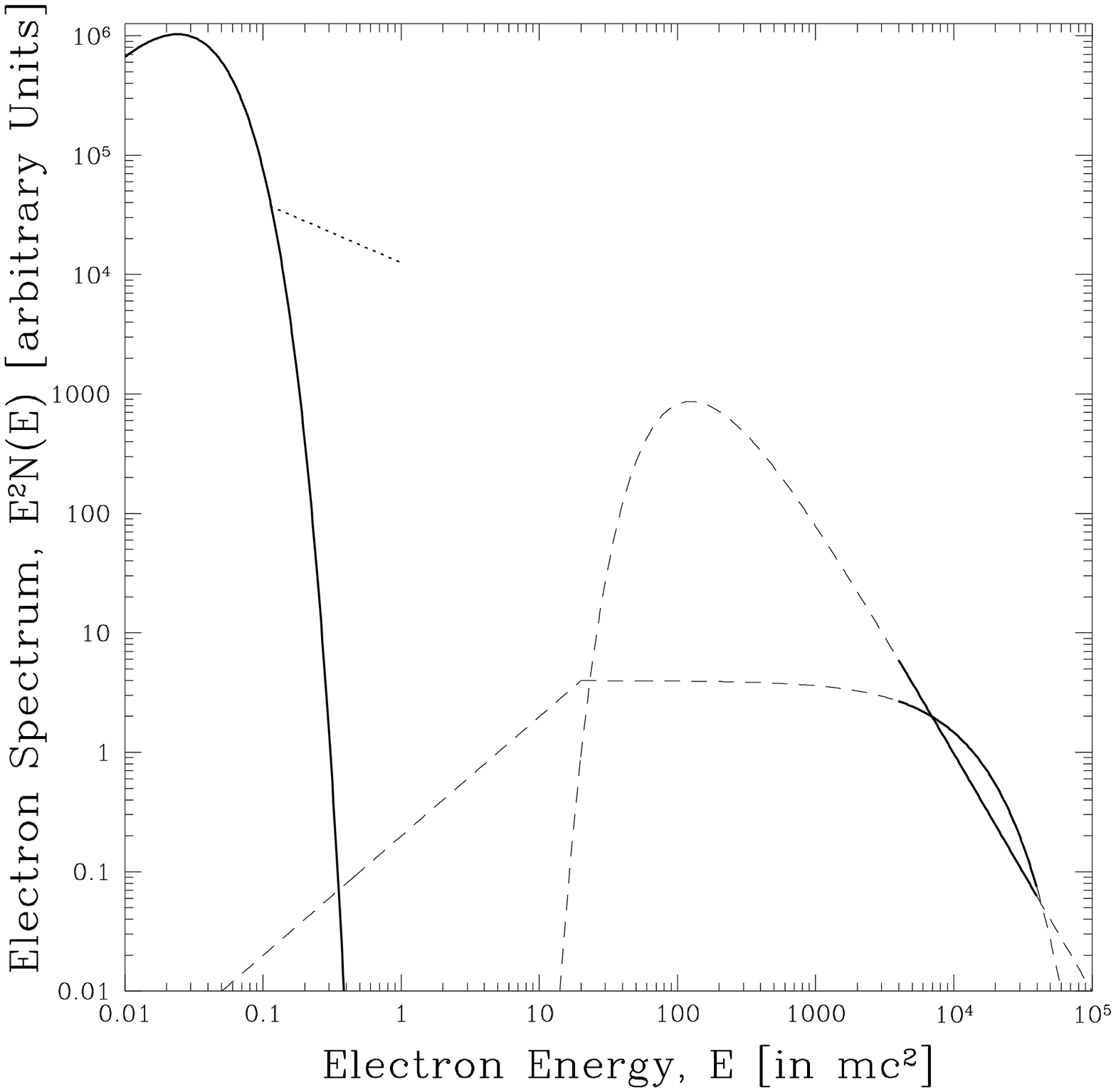}
\caption{\footnotesize {\bf Left Panel:}  Schematic presentation of the $\nu
f(\nu)$ flux of the electromagnetic fields in the ICM of Coma cluster including
the $B$ field.  The two short dashed lines show two different fits to the radio
data. {\bf Right Panel:}  Schematic spectra of the thermal ($T=10^8$ K) and two
nonthermal electrons responsible for the radio emission (solid lines).  The
dashed lines show maximal extrapolations of spectra so that one avoids
unacceptably high rate of heating of the ICM plasma.  The dotted line is the
electron spectrum for the NTB model.  This clearly exceeds the heating limit.
} 
\end{figure}

\section{Radiation Mechanisms}

The HXR emission could be produced via IC scattering of CMB photons by the same
population of relativistic electrons responsible for the radio emission (see
e.g.  Sarazin \& Lieu 1998) shown by the solid lines in Figure 2 (right panel).
However, simple arguments show that this scenario requires a field strength
$B_\perp< 0.2 \mu$G, which is much smaller than values of several $\mu$G
deduced from Faraday rotation mentioned above and equipartition arguments.
Consequently, several workers have proposed nonthermal bremsstrahlung (NTB) for
the HXR emissions (En\ss lin et al.  1999, Sarazin \& Kempner 2000, Blasi
2000).  The dotted line in Figure 2 (right panel) shows the spectrum of the
required electrons.  However, as shown by Petrosian (2001, {\bf P01}), the NTB
process faces a serious difficulty, which is hard to circumvent.  This is
because compared to Coulomb losses the bremsstrahlung yield is very small;
$Y_{\rm br} \sim 3\times10^{-6}(E/25 {\rm keV})^{3/2}$ (see Petrosian 1973).
Thus, for a HXR luminosity of $4\times 10^{43}$ erg s$^{-1}$ (for Coma), a
power of $L_{\rm HXR}/Y_{\rm br} \sim 10^{49}$ erg s$^{-1}$ is fed into the
ICM, increasing its temperature to $T\sim 10^8$ K after $3\times 10^7$ yr or to
$10^{10}$ K in a Hubble time!  Therefore, {\it the NTB emission phase, if any,
must be very short lived}.

This inefficiency of the NTB appears more serious than the inefficiency of the
IC relative to the synchrotron process.  There are several arguments which
indicate that a higher $B$ field can be tolerated in the IC model (see {\bf
P01}).  Briefly, this discrepancy can be alleviated by i) a more realistic
electron spectral distribution (e.g.  Exponential spectral break beyond $\gamma
\sim 10^4$); ii) a non-isotropic pitch angle distribution (Epstein 1973); and
iii) spatial inhomgeneities (Goldschmidt \& Rephaeli 1993, Govoni et al.
2003).  Finally, the Faraday rotation measures may give a somewhat biased view
of the $B$ field by selecting clusters with the highest values of $B$ while the
EUV or HXR observations favor clusters with low values of $B$.  The cluster
RX-J0658 was chosen for observations because it was estimated that it should
have relatively high IC flux of $\sim 7 \times 10^{-12}$ erg cm$^{-2}$ s$^{-1}$
which is what is observed.  This increases our confidence in the validity of
the IC model.

\section{Acceleration Mechanism}

It turns out that the acceleration mechanism of electrons can also be
constrained significantly, even though we have very limited data.  This
mechanism should produce the relativistic electron spectra shown in Figure 2
(right panel).  The lifetimes of these electrons are longer than their crossing
time, $T_{\rm tr}\sim 3\times 10^6$ yr.  Therefore, these electrons will escape
the cluster and radiate most of their energy outside it unless there exists
some scattering agent with a mean free path $\lambda_{\rm scat}\sim 1$ kpc to
trap the electrons in the ICM for at least a timescale of $T_{\rm
esc}=(R/\lambda_{\rm scat})T_{\rm tr}\sim 3\times 10^9$ yr, for cluster size
$R\sim 1$Mpc.  {\bf Turbulence} can be this agent and can play a role in
stochastic acceleration directly, or indirectly in acceleration by {\bf
shocks}.  Both shocks and turbulence can presumably be produced during merger
events.  Several lines of arguments point to an ICM which is highly turbulent.
The possible scenarios of acceleration by turbulence and/or shocks are explored
in {\bf P01} leading to the following conclusions:  i) The seed electrons
cannot be the ICM thermal electrons for the same reason that the NTB fails as a
source of HXRs, namely because it will lead to excessive heating of the ICM.
Therefore, we require injection of high energy ($>50$ MeV) electrons into the
ICM, presumably from galaxies or AGNs.  ii) The short lifetimes of the relevant
electrons with respect to $T_{\rm esc}$ and the small $\lambda_{\rm scat}$
imply a continuous and {\it in situ} acceleration process.  iii) A {\it steady
state} model seems natural but it leads to a flatter spectrum than required
unless the turbulence has an unreasonably steep spectrum.  iv) {\it Time
Dependent Models} can produce the desired spectra but only for a short period
($\sim 3\times 10^8$ to $10^9$ yr) implying an {\it episodic acceleration
process} induced by merger activity.

{\bf Acknowledgment:} This work is partially supported by the NASA grant NAG5 
13031. I would like to thank my collaborators G. Madejski and K. Luli for 
permission to present  our unpublished data here and note that many authors 
have contributed significantly to the subject of this paper whose work I could 
not fully acknowledge because of space limitations.

\end{document}